\documentclass[
 reprint,
 amsmath,amssymb,
 aps,
]{revtex4-2}

\usepackage{graphicx}% Include figure files
\usepackage{bm}% bold math
\usepackage[colorlinks=true, linkcolor=blue,anchorcolor=blue, citecolor=blue,urlcolor=blue]{hyperref}

\usepackage[normalem]{ulem}
\begin{document}
\bibliographystyle{apsrev4-1}
\preprint{APS/123-QED}

\title{Strain-induced frustrated helimagnetism and topological spin textures in LiCrTe$_{2}$}
\author{Weiyi Pan$^{1}$}
\email{pwy20@mails.tsinghua.edu.cn}

\author{Junsheng Feng$^{2,3}$}

%\author{Wenhui Duan$^{1,2,3,4}$}

\affiliation{$^{1}$State Key Laboratory of Low Dimensional Quantum Physics and Department of Physics, Tsinghua University, Beijing, 100084, China\\
%$^{2}$Tencent Quantum Laboratory, Tencent, Shenzhen, Guangdong 518057, China\\
%$^{3}$Institute for Advanced Study, Tsinghua University, Beijing 100084, China \\
%$^{4}$Frontier Science Center for Quantum Information, Beijing, China\\
$^{2}$Key Laboratory of Computational Physical Science (Ministry of Education), State Key Laboratory of Surface Physics, and Department of Physics, Fudan University, Shanghai 200433, China\\
$^{3}$School of Physics and Materials Engineering, Hefei Normal University, Hefei 230601, China}

\date{2023.3}

\begin{abstract}
By performing first-principles calculations in conjunction with Monte Carlo simulations, we systematically investigated the frustrated magnetic states induced by in-plane compressive strain in LiCrTe$_{2}$. Our calculations support that the magnetic ground state of LiCrTe$_{2}$ crystal is A-type antiferromagnetic (AFM), with an in-plane ferromagnetic (FM) state and interlayer AFM coupling. Furthermore, it is found that compressive strain can significantly alter the magnetic interactions, giving rise to a transition from an in-plane FM to an AFM state, undergoing a helimagnetic phase. Remarkably, a highly frustrated helimagnetic state with disordered spin spirals under moderate strain arises from the competition between spiral propagation modes along distinct directions. In addition, various topological spin defects emerge in this frustrated helimagnetic phase, which are assembled from various domain wall units. These topological defects can be further tuned with external magnetic fields. Our calculations not only uncover the origin of exotic frustrated magnetism in triangular lattice magnetic systems, but also offer a promising route to engineer the frustrated and topological magnetic state, which is of significance in both fundamental research and technological applications.

\end{abstract}
\maketitle
\section{Introduction}
The intriguing magnetic phenomena in triangular lattice (TL) magnets, in which each corner of the triangular sites is occupied by a magnetic atom, are a subject of great interest in contemporary condensed matter physics. Specifically, geometric frustration in TL magnets gives rise to strong competition between distinct magnetic interactions\cite{moessner2006geometrical,strong}. Consequently, various exotic magnetic states, such as helical magnetism\cite{TD7,HEEE,HE2,HE3,HE4}, quantum spin liquid\cite{spinliquid1,spinliquid2,spinliquid4}, and topological spin textures\cite{TD5,TD7, TD8,TD9}, could emerge in TL magnets. This makes the TL magnet not only a wonderful playground for novel magnetic phenomena but also a promising platform for advanced electronic and spintronic applications in the future.

As a representative of TL magnets, the family of ACrX$_{2}$ (A = Li, Na, K, Au, Ag, Cu; X = O, S, Se, T) compounds has attracted much attention due to their diversity in magnetic properties\cite{ACX1,ACX2,ACX3,ACX4,ACX5,ACX6,ACX7,ACX8,LCT,LCT2,LCT3,LCT4}. For example, NaCrTe$_{2}$ exhibits in-plane ferromagnetic (FM) order together with interlayer antiferromagnetic (AFM) coupling (named as A-type AFM state)\cite{ACX6,ACX7}, while its isostructural partner, AgCrSe$_{2}$, hosts a helimagnetic state\cite{ACX4}. Moreover, by substituting Te with Se and Na with Li, up-up-down-down spin structures formed in LiCrSe$_{2}$\cite{ACX8}. 
Commonly, the diversity of magnetic states in ACrX$_{2}$ compounds originates from the competition between Heisenberg interactions. Thus, by tuning the strength of the exchange interactions, the above competition would be altered and it is possible to artificially realize multiple novel magnetic phases in ACrX$_{2}$. 
Generally, the strength of the exchange interactions depends on the competition between the direct d-d AFM exchange interaction and the indirect FM superexchange interaction among magnetic ions in an ACrX$_{2}$ compound. More importantly, both direct and indirect exchanges are closely correlated with the structural features of the compound. Hence, it is reasonable to tune the magnetism by modulating the lattice parameters of the compound through applying external strains. However, some issues still remain: How does strain affect the magnetic phases of typical ACrX$_{2}$ systems? Can exotic spin structures such as frustrated magnetism and topological spin defects emerge in typical ACrX$_{2}$ systems under external strain? 

In this work, we focus on the LiCrTe$_{2}$ compound, which has been synthesized in recent experiments\cite{LCT,LCT2}. Noticeably, Kobayashi \textit{et al}. suggested that it possibly possesses a helical magnetic structure through their transport measurements\cite{LCT}, whereas Nocerino and coworkers proposed that it has a trivial A-type AFM state at approximately room temperature according to neutron diffraction measurements\cite{LCT2}. These contradictory results give rise to the possibility that LiCrTe$_{2}$ possesses intrinsically multiple magnetic phases, which could be exhibited under proper external perturbation of its structure. Therefore, it is valuable to investigate how external strain influences the magnetic states of LiCrTe$_{2}$. We utilized first-principles calculations and Monte Carlo (MC) simulations to systematically investigate strain-controllable magnetic states. Our findings indicate that compressive strain sensitively alters the competition of different magnetic interactions in LiCrTe$_{2}$, inducing a transition from an intralayer FM state to an AFM state. Interestingly, on the transfer path from the FM to AFM state, a highly frustrated helimagnetic state with disordered spin spirals emerges. This frustration can be attributed to the competition between two spiral propagation modes along distinct high-symmetric directions. In addition, abundant topological defects are predicted in the frustrated helimagnetic state, and these topological spin textures can be further tuned with an external magnetic field. Our calculations deepen our understanding of the strain-regulating frustrated magnetism in the TL compound.

\section{Methods}

\subsection{First-principles calculations}
We carried out first-principles calculations based on the projector augmented-wave (PAW) method\cite{PAW} implemented in the Vienna \textit{ab initio} simulation package (VASP)\cite{VASP}. During the calculations, the exchange correlation effect was considered within the framework of the generalized gradient approximation (GGA) with the Perdew-Burke-Ernzerhof functional\cite{PBE}. Additionally, we applied the GGA+\textit{U} method\cite{U} to describe the localized d orbitals of Cr atoms, where a small value of \textit{U} = 0.5 eV in the Dudarev implementation was chosen so that the obtained magnetic properties could be consistent with those reported in a previous experiment(see Fig. S2\cite{SM}). The vdW correction at the DFT-D3 level was included so that the dispersion forces could be effectively described\cite{DFT-D3}. We employed the plane-wave basis with an energy cutoff of 400 eV, and the Brillouin zone was sampled by a $15\times15\times4$ $\Gamma$-centered mesh. The convergence criteria for the total force and energy were set to $10^{-3}$ eV/\AA \ and  $10^{-6}$ eV, respectively.

\subsection{Monte Carlo simulations}
Based on the magnetic parameters obtained from first-principles calculations, we performed parallel tempering Monte Carlo (PTMC) simulations\cite{PTMC} as implemented in PASP package\cite{PASP1,PASP2}  to obtain not only the low-temperature magnetic structure , but also the specific heat and susceptibility as a function of temperature. During the PTMC simulation, 150000 MC steps are performed for each configuration. Two kinds of large supercells, $50\times 25 \times 6$ containing 15000 Cr ions and $30\times15\times6$ containing 5400 Cr ions, both of which are based on a rectangular cell defined by $\textbf{a}'$,$\textbf{b}'$ and $\textbf{c}'$, are used. Here, $\textbf{a}' = \textbf{a}$, $\textbf{b}' = \textbf{a}+2\textbf{b}$, $\textbf{c}' = \textbf{c}$, in which \textbf{a}, \textbf{b} and \textbf{c} are the original lattice vectors. To further optimize the spin configurations, conjugate gradient (CG) optimization\cite{CG} is applied after the MC simulations. During the CG optimization, the direction of each spin is locally optimized until the force on each spin is minimized. The criterion of energy convergence of CG calculations is set to $10^{-6}$ eV. By combining the MC simulations with CG optimizations, the resulting spin configurations are all located at energy minima.

\begin{figure*}[ht]
\includegraphics[scale = 0.44 ]{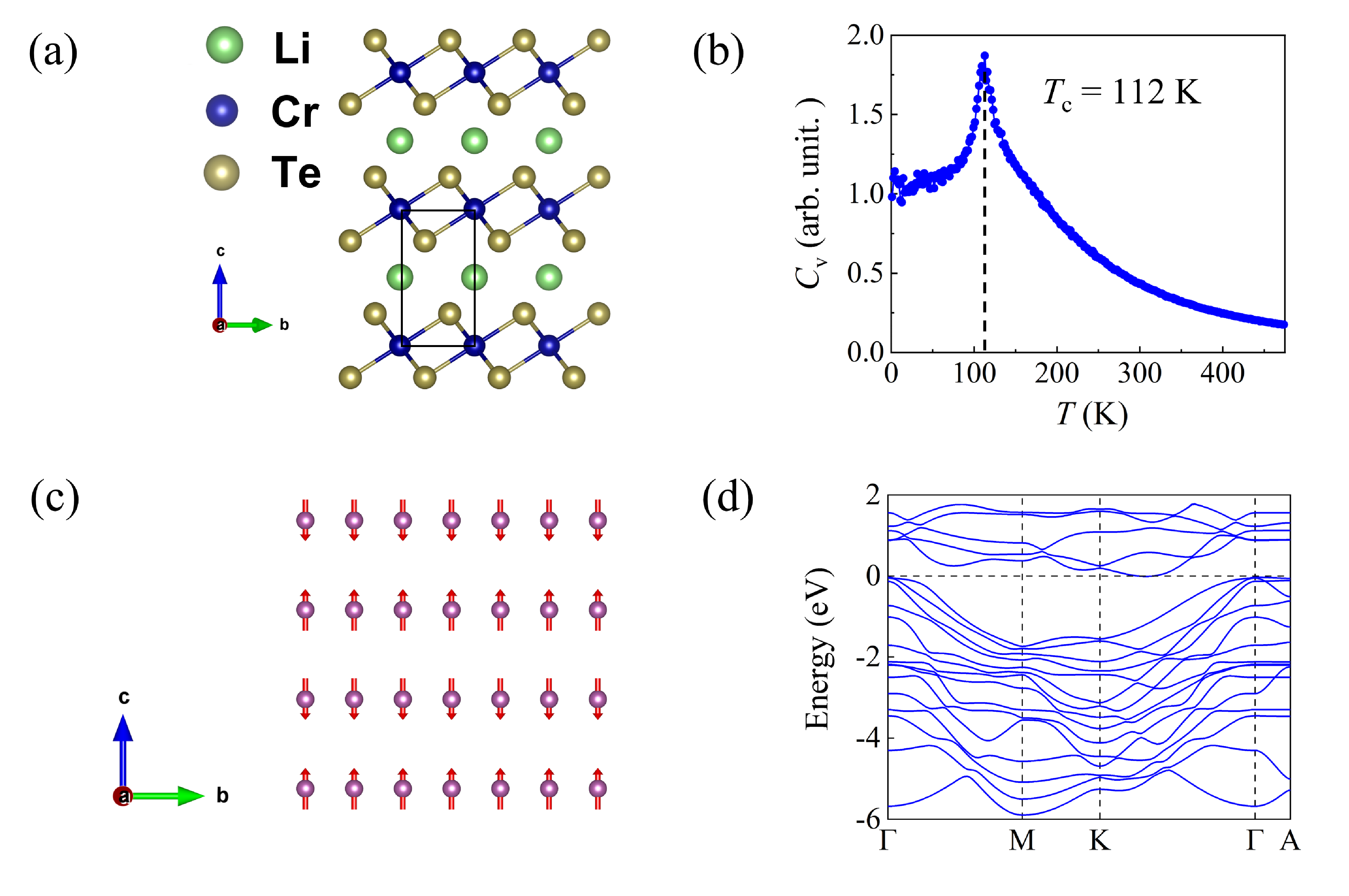}
\caption{\label{1} (a) The crystal structure of LiCrTe$_{2}$.  (b) The electronic band structures of LiCrTe$_{2}$. Note that the magnetic configuration is fixed to be A-type AFM during the band calculation, and the spin-orbital coupling effect is included. (c) Specific heat of LiCrTe$_{2}$ as a function of temperature. $T_{c}$ is extracted from the maximum specific heat. (d) A schematic illustration of the A-type AFM spin configuration. Only the Cr sites are visible. 
}
\end{figure*}

\subsection{Calculation of topological charges}
We calculated the topological charge on the discrete intralayer spin lattice using the formula given below\cite{tpo}:
\begin{equation}
\begin{aligned}
     & Q = \frac{1}{4\pi}\sum_{l}A_{l} \\
     & \textup{cos}(\frac{A_{l}}{2}) = \frac{1+\textbf{m}_{i}\cdot \textbf{m}_{j} + \textbf{m}_{j}\cdot \textbf{m}_{k} + \textbf{m}_{k}\cdot \textbf{m}_{i}   }{  \sqrt{2(1+\textbf{m}_{i}\cdot \textbf{m}_{j}) (1+\textbf{m}_{j}\cdot \textbf{m}_{k}) (1+\textbf{m}_{k}\cdot \textbf{m}_{i})         }           }\\
     & \textup{sign}(A_{l}) = \textup{sign}[\textbf{m}_{i}\cdot(\textbf{m}_{j}\times \textbf{m}_{k})] 
\end{aligned}    
\end{equation}
where \textit{l} runs over all elementary triangles consisting of nearest-neighboring spin sites locating in the same atomic layer, while $A_{l}$ is the solid angle formed by three unit spin vectors $\textbf{m}_{i}$, $\textbf{m}_{j}$ and $\textbf{m}_{k}$ on one elementary triangle (labeled as \textit{l}). Note that the three sites \textit{i}, \textit{j}, and \textit{k} are counted in an anticlockwise manner.

\section{RESULTS AND DISCUSSION}
\subsection{Basic electronic and magnetic properties}

The crystal structure of LiCrTe$_{2}$ is shown in Fig. \ref{1}(a), in which the Li atoms are sandwiched between 1T-CrTe$_{2}$ layers. %The in-plane and out-of-plane lattice constants are calculated to be \textit{a} = 3.97 \AA and \textit{c} = 6.51 \AA, which fit the experimental value (\textit{a} = 3.965 \AA and \textit{c} = 6.55 \AA) well\cite{LCT2}.  
%To mimic the structure feature reported in experiment, the in-plane lattice constants are fixed as \textit{a} = 3.965 \AA, while the out-of-plane lattice constant \textit{c} is fully relaxed until it reaches its optimal value, \textit{c} = 6.51 \AA, which fits the experimental value (\textit{c} = 6.55 \AA) well\cite{LCT2}.%
The band structure of LiCrTe$_{2}$ was calculated, as shown in Fig. \ref{1}(b). It indicates an indirect band gap of 0.05 eV, with the valence band maximum located at Gamma and the conduction band minimum appearing on the $K-\Gamma $ line. In addition, the flat bands along the $\Gamma-A $ path near the Fermi level indicates the absence of interlayer electronic coupling, characterizing the two-dimensional feature of LiCrTe$_{2}$.

To unveil the magnetic properties of LiCrTe$_{2}$, an effective spin Hamiltonian is given as:

\begin{equation}
\begin{aligned}
    & H= J_{1}\sum_{\langle i,j \rangle}\textbf{S}_{i} \cdot \textbf{S}_{j} + J_{2}\sum_{\langle \langle i,j \rangle \rangle}\textbf{S}_{i} \cdot \textbf{S}_{j} \\
    & +J_{3}\sum_{\langle \langle \langle i,j \rangle \rangle \rangle}\textbf{S}_{i} \cdot \textbf{S}_{j} + A \sum_{i} (S_{iz})^{2} + J^{\textup{T}} \sum_{\langle i,j \rangle \bot} \textbf{S}_{i} \cdot \textbf{S}_{j} \label{eq1}
\end{aligned}
\end{equation}
Here, $J_{1}$, $J_{2}$ and $J_{3}$ denote the first-nearest neighbor (1NN), second-nearest neighbor (2NN), and third-nearest neighbor (3NN) intralayer Heisenberg exchange parameters, respectively. $A$ represents the single-ion anisotropy (SIA) energy and $J^{\textup{T}}$ represents the nearest neighbor interlayer coupling constant.

To achieve the intralayer coupling constants ($J_{1}$, $J_{2}$ and $J_{3}$), we constructed four different intralayer magnetic configurations, such as FM, Stripy, Zigzag-1 and Zigzag-2 (see Fig. S1\cite{SM}). In our treatment, the interlayer coupling is fixed to be FM, and the total energies of the four configurations are given as: 

\begin{equation}
    \begin{aligned}
    & E_{\textup{FM}} = E_{0}+24J_{1}|\textbf{S}|^{2}+24J_{2}|\textbf{S}|^{2}+24J_{3}|\textbf{S}|^{2} \\
    & E_{\textup{Stripy}} = E_{0}-8J_{1}|\textbf{S}|^{2}-8J_{2}|\textbf{S}|^{2}+24J_{3}|\textbf{S}|^{2} \\
    & E_{\textup{Zigzag-1/2}} = E_{0}\pm 8J_{1}|\textbf{S}|^{2}\mp8J_{2}|\textbf{S}|^{2} - 8J_{3}|\textbf{S}|^{2} \\
    \end{aligned}
\end{equation}

By solving the four equations above, the values of parameters ($J_{1}$, $J_{2}$ and $J_{3}$) are achieved.

The value of interlayer coupling parameter $J^{\textup{T}}$ is evaluated from the energy difference between an interlayer FM and an interlayer AFM state (see Fig. S1\cite{SM}), namely,

\begin{equation}
    J^{\textup{T}} = (E_{\textup{FM}}-E_{\textup{AFM}})/4
\end{equation}.

Here, $E_{\textup{FM}}$ ($E_{\textup{AFM}}$) represents the energy of the interlayer FM (AFM) state. %In this case, the intralayer states are fixed to be FM states.
The value of SIA is obtained by calculating the total energy difference between the out-of-plane FM configuration ($E_{z}$) and the in-plane FM configuration ($E_{x}$), namely:

\begin{equation}
    A = E_{z} - E_{x}
\end{equation}

Here, the positive (negative) value of $A$ is an indicator of in-plane (out-of-plane) magnetization. Spin-orbital coupling (SOC) is switched on during the calculation of SIA since SIA is a second-order SOC effect\cite{SIA}.

\begin{table}[ht]
\caption{The values of magnetic parameters of LiCrTe$_{2}$ based on equation (\ref{eq1}). For simplification, we normalized the spins to S = 1 in this work. The energy unit is meV/Cr.}
\begin{ruledtabular}
    \begin{tabular}{ccccc}
     $J_{1}$ & $J_{2}$ & $J_{3}$ & $A$ & $J^{\textup{T}}$ \\
    \hline
     -15.10 & 0.39 & 3.01 & -0.69 & 0.61 \\
    \end{tabular}
\end{ruledtabular}
    \label{tb1}
\end{table}

\begin{figure*}[ht]
\includegraphics[scale = 0.42 ]{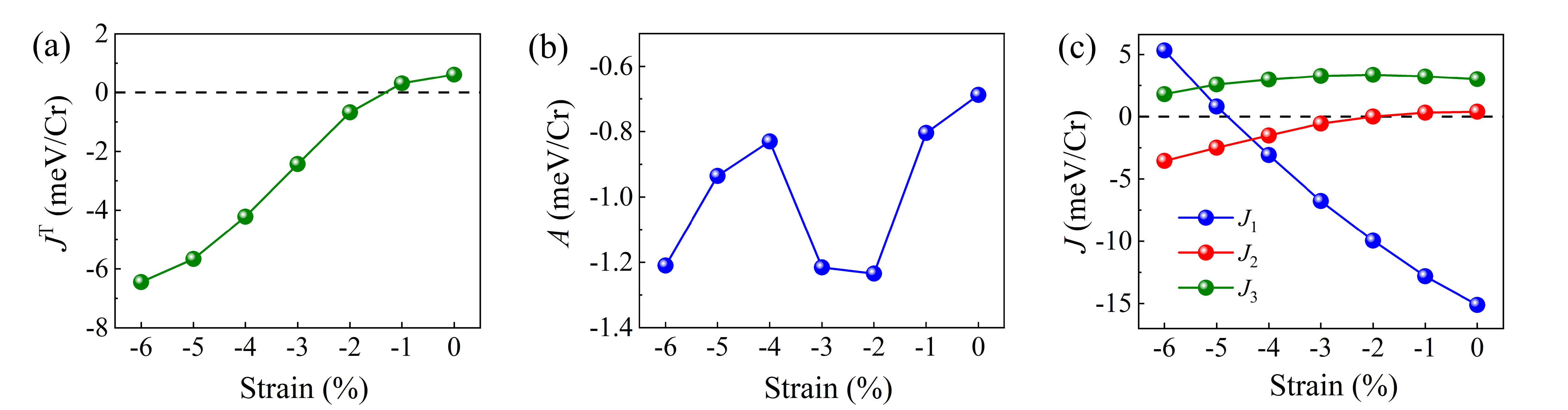}
\caption{\label{2} (a) The interlayer coupling $J^{\textup{T}}$, (b) the SIA, and (c) the intralayer coupling constants ($J_{1}$, $J_{2}$, $J_{3}$) as a function of the in-plane compressive strain.} 
\end{figure*}

The values of all magnetic parameters concerned above are tabulated in Table \ref{tb1}. 
It can be seen that among the parameters of $J_{1}$, $J_{2}$ and $J_{3}$, the amplitude of $J_{1}$ is the largest, and that of $J_{2}$ is the smallest. The amplitude of $J_{3}$ is one order of magnitude larger than that of $J_{2}$. Clearly, the dominant exchange interaction is the FM coupling between the nearest neighbors ($J_{1}$). Meanwhile, the AFM exchange interaction between the third nearest neighbors ($J_{3}$) is nonnegligible. So, among various couplings  between $J_{1}$, $J_{2}$ and $J_{3}$, the coupling between $J_{1}$ and $J_{3}$ is the main tone. In addition, the value of SIA is a negative value, indicating that the easy magnetization axis lies in the out-of-plane direction. The parameter of $J^{\textup{T}}$ is positive, implying an AFM feature occurring in the interlayer coupling.

Since we have the parameters in the spin Hamiltonian (\ref{eq1}), we then carried out MC simulations and obtained the specific heat as a function of temperature, as shown in Fig. \ref{1}(c). The location of the peak corresponds to a critical temperature $T_{c}$ = 112 K, which agrees with the value of 125 K from neutron diffraction experiment\cite{LCT2}. Furthermore, the magnetic ground state obtained by MC simulation is an A-type AFM state with intralayer FM state and interlayer AFM coupling (as shown in Fig. \ref{1}(d)), which is consistent with the previous neutron diffraction experimental report as well\cite{LCT2}. Till now, based on our proposed effective spin Hamiltonian described in formula (\ref{eq1}), we successfully reproduced the key experimental phenomena reported by Nocerino \textit{et al}\cite{LCT2}.  This strongly supports that our proposed effective spin Hamiltonian, together with the determined parameters listed in Table \ref{tb1}, is robust.

\subsection{Effect of in-plane strain on magnetic interactions and critical temperatures}

\begin{figure*}[t]
\includegraphics[scale = 0.42 ]{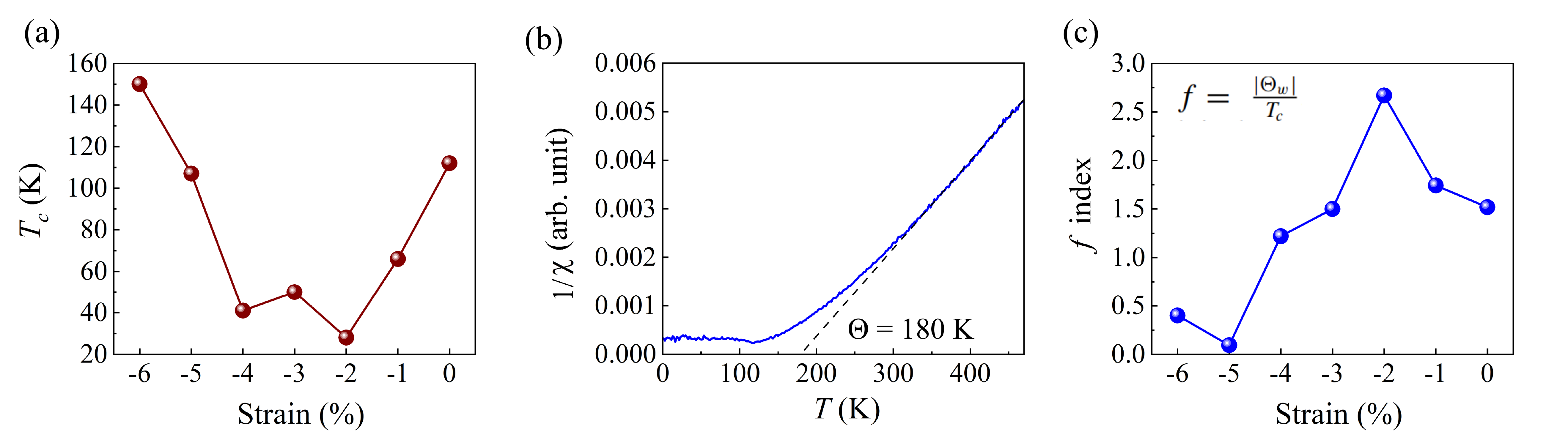}
\caption{\label{3} (a) The critical temperature ($T_{\textup{c}}$) as a function of strain. (b) Inverse of the susceptibility as a function of temperature without strain. (c) Frustration index $f$ as a function of strain.} 
\end{figure*}

As mentioned in introduction, Kobayashi \textit{et al}., according to their experiment, speculated that LiCrTe$_{2}$ is likely to be in a helimagnetic state\cite{LCT}. This, however, is not exhibited in our above calculations where the magnetic ground state of LiCrTe$_{2}$ is an A-type AFM state. As being reported before, there is a strong coupling between spin and lattice in chromic chalcogenides\cite{SMC1,SMC2,ACX2,ACX3}. Hence, external perturbation such as straining of the lattice can lead to corresponding changes in the magnetic structure of the system. Inspired by this feature, we speculate that the proposed helimagnetic state in LiCrTe$_{2}$ probably correlates to the strain of the system to some extent. We, therefore, turn to investigate the strain effect on the magnetism of LiCrTe$_{2}$.

The magnetic parameters of Eq. (\ref{eq1}) as a function of in-plane compressive strain were carefully treated, which is shown in Fig. \ref{2}. It can be seen from Fig. \ref{2}(a) that with larger compressive strain, the interlayer coupling $J^{\textup{T}}$ decreases monotonically from positive value to negative value. This just corresponds to the switching from a weak AFM interlayer coupling to a nonnegligible FM interlayer coupling. Meanwhile, the value of SIA remains negative in the strain range between 0$\%$ and -6$\%$, as shown in Fig. \ref{2}(b), indicating that the easy magnetization axis is robust along the $z$ direction under compressive strain we considered. In particular, the 1NN intralayer Heisenberg interaction $J_{1}$ strikingly increases from negative to positive values with increasing compressive strain, corresponding to the switching from FM coupling to AFM coupling. This would possibly give rise to the transition of the magnetic phase under proper compressive strains. Unlike the case of $J_{1}$, both the 2NN interaction $J_{2}$ and the 3NN interaction $J_{3}$ gently change with the loaded compressive strain. Overall, under different compressive strains, the values of $J_{1}$, $J_{2}$ and $J_{3}$ in the system are different. This implies that there exists different competition between FM and AFM under different compressive strain. 

Basically, the competition between distinct magnetic couplings, which are tightly correlated with external compressive strains, would affect the magnetic phase transition temperatures. We thus study the relationship between critical temperature $T_{\textup{c}}$ associated with the magnetic phase transition and compressive strain for the concerned system. From our calculations, we found that the value of $T_{\textup{c}}$ non-monotonically changes in the strain range we considered, as shown in Fig. \ref{3}(a). Of this considered strain, $T_{\textup{c}}$ is significantly suppressed within a strain range of -2$\%$ to -4$\%$. Remarkably, $T_{\textup{c}}$ reaches its minimum value under a -2$\%$ strain, which corresponds to the most significant suppression of the ordered magnetic state. Nevertheless, once the compressive strain reaches -5$\%$, the value of $T_{\textup{c}}$ rebounds significantly, which is even comparable with that of the FM state without strain. These results suggest that the critical temperature could be significantly tuned by external strain.

To investigate whether compressive strain induces frustrated magnetic states, we calculated the frustration index $f$, which can reveal the degree of spin frustration, as a function of strain. Here the frustration index is defined as $f$ = $\frac{|\Theta_{w}|}{T_{\textup{c}}}$, with $\Theta_{w}$ being the Curie‒Weiss temperature and $T_{\textup{c}}$ being the critical temperature. Here the value of $\Theta_{w}$ can be determined by linearly interpolating the inverse of susceptibility down to zero, as guided with dashed line in Fig. \ref{3}(b). Based on this definition, $f \approx 1$ corresponds to the case of nonfrustrated FM state, and $f \textgreater 1$ suggests the existence of magnetic frustration\cite{FR3}. As shown in Fig. \ref{3}(c), the frustration index $f$ reaches its maximum value under -2$\%$ strain. This implies that the frustrated magnetic state could form under compressive strain between -2$\%$ and -4$\%$, particularly at -2$\%$ strain. Note that the $f$ index is strongly suppressed by loading compressive strain larger than -5$\%$, signaling the emergence of magnetic ordered states.

\begin{figure*}[t]
\includegraphics[scale = 0.48]
{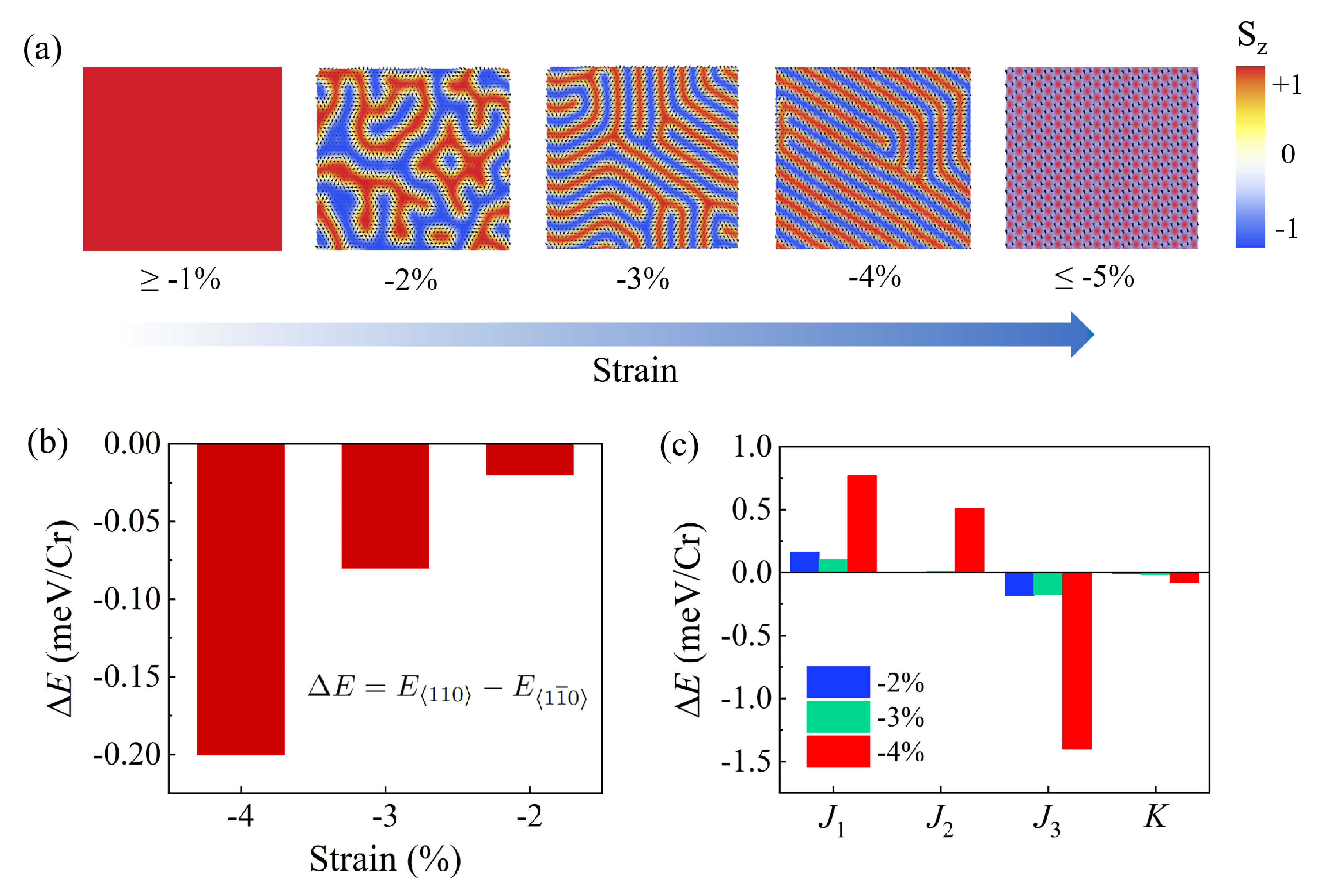}

\caption{\label{4} (a) Intralayer magnetic structures of LiCrTe$_{2}$ under different compressive strains. The color map indicates the out-of-plane spin component. Note that when the compressive strain is smaller than -1$\%$, the magnetic state in each layer is a trivial FM state. Moreover, under compressive strain stronger than -2$\%$, all layers have the same spin textures due to the interlayer FM coupling. Thus, we only show the magnetic configuration of a single layer. The (b) total energy difference and (c) interaction decomposed energy difference between propagation mode along $\langle 110 \rangle$ and $\langle 1 \overline{1 } 0 \rangle$ are also shown. The energy difference is defined as $\Delta E = E_{\langle 110 \rangle} - E_{\langle 1 \overline {1 } 0 \rangle}$. Note that the interlayer coupling does not contribute to the energy difference, so it is excluded from (c).} 
\end{figure*}

\subsection{Strain-induced magnetic phase transition and frustrated magnetism}

To visually display the magnetic states under different strain levels, we performed MC simulations on large supercells with spin Hamiltonian described in equation (\ref{eq1}) in which the parameters are strain-dependent, and the results are presented in Fig. \ref{4} (a). Apparently, when the loaded compressive strain is no more than -1$\%$, the intralayer magnetic configuration remains FM; Enhancing the compressive strain to be -2$\%$, a wormlike helimagnetic texture with winding domains emerges. In this case, the spin texture is disordered, which is consistent with the prior speculation based on Fig. \ref{3}(c) in which a frustrated magnetic state would emerge under -2$\%$ compressive strain. It is noted that such irregular domains are always present in the system at strain of -2$\%$, for they appear in several MC simulations we conducted. Thus, this disordered helimagnetic state is not a result of randomness of the MC simulations but stems from other reasons, which will be discussed later in this paper. By further increasing the compressive strain level up to -3$\%$, the spin texture evolves into a more ordered state. Specifically, the domains become narrowed and their boundaries become straightened. Meanwhile, one can clearly observe domains that propagate along three directions such as the $\langle 110 \rangle$ or the equivalent $\langle 100 \rangle$ and $\langle 010 \rangle$ directions, although some local winding domains are still visible (see Fig. S4 in SM\cite{SM}). Moreover, the domains could be further narrowed and their boundaries becomes more straightened with increasing compressive strain up to -4$\%$. Finally, an ordered AFM configuration appears when the compressive strain reaches -5$\%$. The occurrence of an ordered AFM state is consistent with the prior speculation based on Fig. \ref{3}(c). In total, when the applied strain on the system is proper, the helimagnetic texture proposed by Kobayashi \textit{et al}.\cite{LCT} appears surely. Therefore, our suggested effective spin Hamiltonian described in formula (\ref{eq1}) together with the strain-dependent parameters can unite the distinct results reported in different experiments \cite{LCT,LCT2}.

\begin{figure*}[t]
\includegraphics[scale = 0.48]
{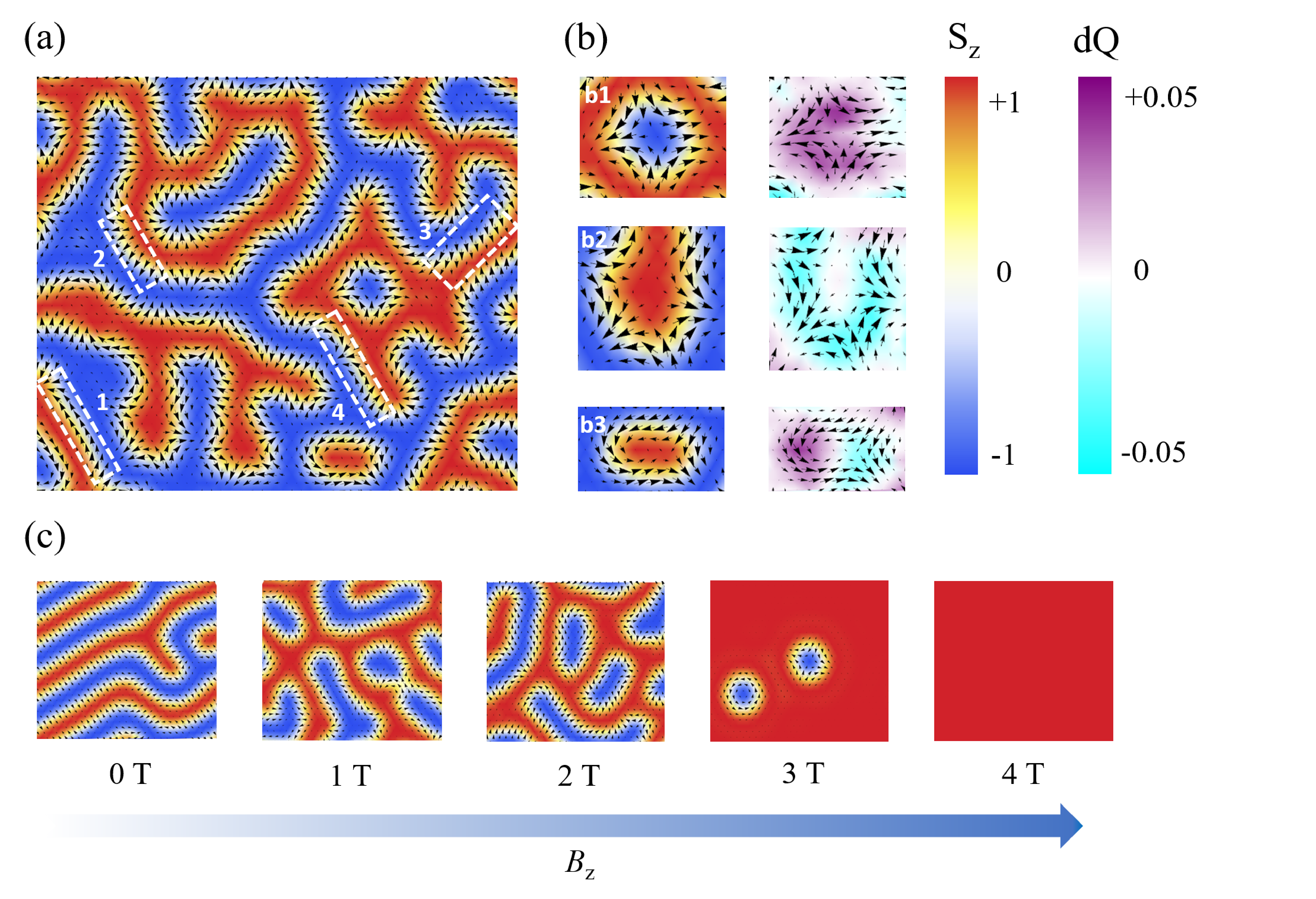}

\caption{\label{5}  (a) A close view of the intralayer magnetic configuration of LiCrTe$_{2}$ under -2$\%$ compressive strain. The (1) Bloch domain wall, (2) N\'eel domain wall, (3) mixed domain wall, and (4) Bloch line are circled with dashed lines, respectivley. (b) Snapshots of topological spin textures, including (b1) an antiskyrmion ($Q$ = 1), (b2) a high-$Q$ antiskyrmion ($Q$ = -2), and (b3) a skyrmion-antiskyrmion pair ($Q$ = 0). Their local topological charge distributions are also shown. The full distribution of topological charge can be seen in Fig. S7 in SM \cite{SM}. (c) The evolution of the intralayer magnetic configuration under magnetic fields with a -2$\%$ compressive strain. The red‒blue and purple-cyan color maps indicate the out-of-plane spin component and local topological charge, respectively.} 
\end{figure*}

Since the values of $J_{1}$, $J_{2}$ and $J_{3}$ as a function of compressive strain level have been achieved above, we, extendedly, predicted the magnetic phase diagram for the classical spin ground state by using the Fressier method\cite{FRE}. The energy $E(\textit{\textbf{k}})$ as a function of magnetic propagation vector $\textit{\textbf{k}}$ is given, as shown in Fig. S5\cite{SM}, and the classical spin ground state resulting from the Fressier method can be viewed as a helimagnetic state with a propagation vector $\textit{\textbf{k}}_{0}$, where $\textit{\textbf{k}}_{0}$ minimizes the energy $E(\textit{\textbf{k}})$. It can be clearly seen that for small compressive strain (no larger than -1$\%$), the global energy minimum is located at the $\Gamma$ point, implying that the classical ground state is FM. Nevertheless, in the strain range between -2$\%$ and -4$\%$, the global minimum appears along the $K-\Gamma$ path, indicating a helimagnetic state propagating along the $\langle 110 \rangle$ direction (as well as its equivalent $\langle 100 \rangle$ and $\langle 010 \rangle$ directions, which will not be addressed in the following text). Note that there is another local minimum lying on the $ \Gamma-M$ path, corresponding to a metastable helimagnetic state propagating along the $\langle 1 \overline{1 } 0 \rangle$ direction (as well as its equivalent $\langle 120 \rangle$ and $\langle 210 \rangle$  directions, which will not be addressed in the following text). These two local minima have comparable energies, and thus the two spiral propagation modes along $\langle 110 \rangle$ and $\langle 1 \overline{1 } 0 \rangle$ coexist in the system. When the compressive strain reaches -5$\%$, the global energy minimum shifts to the $K$ point, which is located at the corner of the 1st Brillouin Zone, belonging to an AFM classical ground state. Based on the above analysis, we plot a phase diagram as a function of strain (as shown in Fig. S6\cite{SM}). It is evident from the phase diagram that as the system undergoes compressive strain, its magnetic configuration transfers from an FM phase to an AFM phase, passing through helimagnetic phases in between. These findings are in good agreement with the results of MC simulations shown in Fig. \ref{4}(a).

Now, we turn to understand why the spin spirals are disordered and frustrated at -2$\%$ compressive strain, and become more narrowed and ordered with increasing compressive strain level. We argue that the competition between two coexisting spiral propagation modes along $\langle 110 \rangle$ and $\langle 1 \overline{1 } 0 \rangle$ as revealed above leads to the spin frustration. To assess this, it is necessary to investigate the spiral propagation modes along $\langle 110 \rangle$ and $\langle 1 \overline{1 } 0 \rangle$. 
Based on our MC simulations followed by CG optimization, the propagation periodic length of spin spirals along $\langle 110 \rangle$ ($\langle 1 \overline{1 } 0 \rangle$) are determined to be 7.25 $a$ (7.36 $a$) at -2$\%$ strain, 5.8 $a$ (5.84 $a$  ) at -3$\%$ strain, and 4.83 $a$ (5.39 $a$) at -4$\%$ strain, with $a$ = 4.04 \AA \ being the in-plane lattice constant. Here, with increasing compressive strain, the decrement of the periodic length matches the narrowing domains shown in Fig. \ref{4}(a). Next, we calculated the energies of spiral propagation modes along $\langle 110 \rangle$ and $\langle 1 \overline{1 } 0 \rangle$ and evaluated their difference, which are shown in Fig. \ref{4}(b). Clearly, the energy of the spiral model along the $\langle 110 \rangle$ direction is lower than that along the $\langle 1 \overline{1 } 0 \rangle$ direction at each case of considered strains. So, the ground state spiral mode always prefers the $\langle 110 \rangle$ direction in the strained system. Especially, when the loaded compressive strain is -2$\%$, the energy difference between the two spiral modes is quite small, and thus these two modes could coexist in the system, giving rise to a highly directionless frustrated magnetic state. With a larger compressive strain, the magnitude of the energy difference increases. In this case, the spiral mode along $\langle 110 \rangle$ dominates, while that along $\langle 1 \overline{1 } 0 \rangle$ gradually vanishes, forcing the boundaries of the magnetic domains to be straightened, and exhibiting the ordered magnetic configuration.   

Furthermore, we decomposed the energy difference between spiral propagation modes along the $\langle 110 \rangle$ and $\langle 1 \overline{1 } 0 \rangle$ directions onto different magnetic parameters based on the spin Hamiltonian (\ref{eq1}), as shown in Fig. \ref{4}(c). As can be seen, the energy difference is mainly contributed by $J_{1}$ and $J_{3}$. Specifically, the energy difference contributed by $J_{3}$ is negative, while that by $J_{1}$ is positive. This means $J_{1}$ prefers a propagation along $\langle 1 \overline{1 } 0 \rangle$ direction, while $J_{3}$ favors $\langle 110 \rangle$ direction, which is a consequence of $J_{1}$-$J_{3}$ competition. With a larger compressive strain, the magnitude of the decomposed energy difference from $J_{3}$ greatly increases, which is responsible for the spiral propagation modes along $\langle 110 \rangle$ being more dominant. Such results not only conform our above-mentioned speculation where the competition between spiral propagation modes along the $\langle 110 \rangle$ and $\langle 1 \overline{1 } 0 \rangle$ directions would give rise to a highly disordered magnetic state, but also imply that the extent of disorder depends on the magnitude of the energy difference between those two modes.

\subsection{Topological spin defects in the frustrated magnetic state}

Frustrated magnetic states are commonly believed to be the playground for exotic magnetic phenomena. Since the magnetic state becomes highly frustrated under -2$\%$ compressive strain, we wonder whether there are any exotic spin structures lying in this frustrated state. By carefully examining the details of the intralayer magnetic state at -2$\%$, four kinds of coexisting domain wall units, as shown in Fig. \ref{5}(a), are observed: (1) Bloch walls, in which the in-plane spin components are parallel to the domain walls; (2) N\'eel walls, in which the in-plane spin components are perpendicular to the domain walls; (3) Mixed walls, which possess tilted in-plane spin components relative to the domain walls; and (4) Bloch line structures, in which the spins rotate in the $x-y$ plane. Note that different types of domain wall units always emerge together, and they assemble plenty of topological spin defects, some of which are shown in Fig. \ref{5}(b). As shown in Fig. \ref{5}(b1), there is an isolated antiskyrmion with $Q$ = 1, where the spins lying around it rotate in clockwise manner through $2\pi$. This antiskyrmion can be regarded as being constituted by Bloch lines. In addition to common antiskyrmions, the Bloch lines could also assemble antiskyrmions with $Q$ = -2, where spins rotate $-4\pi$ on the path circling the core of such a high-$Q$ antiskyrmion, as shown in Fig. \ref{5}(b2). A skyrmion-antiskyrmion pair with $Q$ = 0 is also visible, as shown in Fig. \ref{5}(b3), which is formed by connecting a Bloch wall with a Bloch line together locally and winding it into a closed curve. We emphasize that the topological defects shown above are all purely induced by frustration instead of extra novel interactions such as the Dzyaloshinskii-Moriya interaction\cite{NoDMI,E3,LCT4,XC2} or higher-order interactions\cite{3Q,Novel3,Novel5}. There might be more exotic topological defects caused by frustration, which will be studied in the future.

Since wormlike domain wall states are commonly sensitive to magnetic fields, it is necessary to study the evolutionary behavior of magnetic states under external fields. For this purpose, we performed MC simulations and included different magnetic fields (labeled $B_{\textup{z}}$) by adding a Zeeman term $H_{z} = g\mu _{B}S_{z}B_{z}$, where $g$ is the Land\'e factor and $\mu _{B}$ represents the Bohr magneton. As shown in Fig. \ref{5}(c), by applying $B_{\textup{z}}$ up to 2 T, the wormlike domain walls form closed cycles, leading to the formation of isolated strip-like spin-down islands. These islands consist of connected skymion-antiskyrmion pairs. With $B_{\textup{z}}$ = 3 T, the skymion-antiskyrmion pairs break apart, which brings about the emergence of isolated skymions and antiskymions in the FM background. Finally, when $B_{\textup{z}}$ reaches 4 T, the skyrmions and antiskyrmions are all eliminated, and the system arrives at the FM phase. Such results imply that it is an ideal approach to induce isolated topological defects in the frustrated magnetic state with a magnetic field, while these isolated topological defects might be useful for future electronic applications.

\section{Conclusion}
In conclusion, first-principles calculations in conjunction with MC simulations provide insights into the spin couplings in LiCrTe$_{2}$. Our calculations show that the magnetic ground state of LiCrTe$_{2}$ is an A-type AFM with intralayer FM state and interlayer AFM coupling. Furthermore, the impact of in-plane compressive strain on the magnetism of LiCrTe$_{2}$ was revealed. This leads to our prediction that strain can significantly alter the magnetic interactions, giving rise to a transition from intralayer FM to intralayer AFM, linked by a helimagnetic phase. Remarkably, the system can exhibit a frustrated helimagnetic state under a moderate strain level, which arises from the competition between different spin spiral propagation modes. In addition, colorful topological spin defects, which are assembled by many domain wall units, are predicted to exist in the frustrated helimagnetic phase. These topological spin defects can be tuned with advantage by applying an external magnetic field. These results not only shed light on the origin and behavior of frustrated magnetic states in realistic systems hosting TL lattices, but also offer a promising avenue to induce and engineer such states.

\section{Acknowledgments}
The authors sincerely thanks Professor Wenhui Duan, Professor Yong Xu, and Professor Changsong Xu for helpful discussions. J.S. Feng  acknowledges the support from Anhui Provincial
Natural Science Foundation (Grant No. 1908085MA10) and
the Opening Foundation of State Key Laboratory of Surface
Physics Fudan University (Grant No. KF2019\_07).

\bibliographystyle{apsrev4-1}

\end{document}